# Optimization of hyperparameters for SMS reconstruction

L. Tugan Muftuler PhD[1], Volkan Emre Arpinar PhD[2,3], Kevin Koch PhD[2,3], Sampada Bhave PhD[2,3], Baolian Yang PhD[4], Sivaram Kaushik PhD[4], Suchandrima Banerjee PhD[5], Andrew Nencka PhD[2,3]

[1]Department of Neurosurgery, Medical College of Wisconsin, 8701 Watertown Plk Rd. Milwaukee, WI, 53226, USA

[2]Department of Radiology, Medical College of Wisconsin, 8701 Watertown Plk Rd. Milwaukee, WI, 53226, USA

[3]Center for Imaging Research, Medical College of Wisconsin, 8701 Watertown Plk Rd. Milwaukee, WI, 53226, USA

[4]GE Healthcare, 3000 N Grandview Blvd. Waukesha, WI, 53188, USA

[5]GE Healthcare, Menlo Park, CA 94025, United States

Corresponding author:

L. Tugan Muftuler, PhD
Associate Professor of Neurosurgery
Medical College of Wisconsin.
8701 Watertown Plank Road
Milwaukee, WI 53226, USA
Email: lmuftuler@mcw.edu
Phone: +1(414) 955-7627

ABSTRACT

PURPOSE

Simultaneous multi-slice (SMS) imaging accelerates MRI data acquisition by exciting multiple image slices with a single radiofrequency pulse. Overlapping slices encoded in acquired signal are separated using a mathematical model, which requires estimation of image reconstruction kernels using calibration data. Several parameters used in SMS reconstruction impact the quality and fidelity of final images. Therefore, finding an optimal set of reconstruction parameters is critical to ensure that accelerated acquisition does not significantly degrade resulting image quality.

METHODS

Gradient-echo echo planar imaging data were acquired with a range of SMS acceleration factors from a cohort of five volunteers with no known neurological pathology. Images were collected using two available phased-array head coils (a 48-channel array and a reduced diameter 32-channel array) that support SMS. Data from these coils were identically reconstructed offline using a range of coil compression factors and reconstruction kernel parameters. A hybrid space (k-x), externally-calibrated coil-by-coil slice unaliasing approach was used for image reconstruction. The image quality of the resulting reconstructed SMS images was assessed by evaluating correlations with identical echo-planar reference data acquired without SMS. A finger tapping functional MRI (fMRI) experiment was also performed and group analysis results were compared between data sets reconstructed with different coil compression levels.

RESULTS

Between the two RF coils tested in this study, the 32-channel coil with smaller dimensions clearly outperformed the larger 48-channel coil in our experiments. Generally, a large calibration region (144-192 samples) and small kernel sizes (2-4 samples) in k y direction improved image quality. Use of regularization in the kernel fitting procedure had a notable impact on the fidelity of reconstructed images and a regularization value 0.0001 provided good image quality. With optimal selection of other hyperparameters in the hybrid space SMS unaliasing algorithm, coil compression caused small reduction in correlation between single-band and SMS unaliased images. Similarly, group analysis of fMRI results did not show a significant influence of coil compression on resulting image quality.

CONCLUSIONS

This study demonstrated that the hyperparameters used in SMS reconstruction need to be fine-tuned once the experimental factors such as the RF receive coil and SMS factor have been determined. A cursory evaluation of SMS reconstruction hyperparameter values is therefore recommended before conducting a full-scale quantitative study using SMS technologies.

## 1. INTRODUCTION

Simultaneous multi-slice (SMS) acceleration is a cornerstone technology of the Human Connectome Project (HCP) [1]. SMS has helped advance neuroimaging research by accelerating the acquisition of high spatial and temporal resolution images with echo planar imaging (EPI). SMS imaging accelerates data acquisition by exciting multiple slices simultaneously. Overlapping slices are then separated using principles similar to conventional parallel imaging techniques [2,3], whereby the spatial variation of receiver coil sensitivities provides additional degrees of freedom to solve a system of equations. However, SMS has an advantage over conventional in-plane acceleration methods for EPI acquisitions. While in-plane acceleration factors only reduce the duration of the k-space readout, SMS acceleration also reduces the overhead of spin excitation and preparation for individual slices.

Since its initial inception, substantial work has been devoted to the development and refinement of the SMS technique, particularly for EPI, which led to major improvements in data acquisition [4,5] and image reconstruction [5,6]. Separating the aliased voxels to yield unaliased individual slices utilizes data fitting procedures based on calibration data. Such data fitting algorithms require selection of a set of hyperparameters for robust convergence.

Several groups have studied the parameters that affect in-plane GRAPPA acceleration. For instance, Bauer et al analyzed the effects of auto-calibrating signal (ACS) lines, reconstruction kernel size and acceleration factor and reported a set of optimal values for typical MRI exams [7]. Huang and Duensing formulated the kernel truncation and matrix inversion errors in GRAPPA and demonstrated how they are related to coil sensitivity maps and kernel dimensions [8]. Similarly, the tradeoff between fit accuracy and stability in GRAPPA reconstruction was studied by Nana et al in order to find the optimal kernel support [9]. They concluded that if the kernel size was too small, it would not capture the complexity of the data. On the other hand, if it was too large, it became too sensitive to noise. However, similar detailed analyses of hyperparameter choices have not yet been performed for SMS techniques.

In addition to standard calibration and kernel parameters, regularization is also widely utilized for in-plane GRAPPA reconstruction. Regularization of numerical optimization enforces stability in kernel coefficient estimation, which can reduce noise and yield more robust image reconstructions [10]. However, the effectiveness of regularization approaches depends on the amount and method of regularization. Optimal regularization weights for in-plane GRAPPA reconstruction were studied by Qu et al [11]. They compared regularization weights estimated by different methods and concluded that the discrepancy principle provided the best calibration kernel estimates.

Another tool that gained popularity for reconstruction of parallel imaging data is coil compression [12–16]. Wide use of receive coils with high channel count led to an increase in computational burden for advanced image reconstructions. Coil compression reduces the computational load by generating a smaller number of virtual coils (VC) with maximally orthogonal sensitivity profiles [12,15,16]. Since there are significant spatial correlations between RF coil sensitivity profiles in most coil arrays, there is redundant information coming from the receiver channels. This redundancy can be minimized by estimating a set of virtual coil channels that form a smaller set of orthogonal basis functions. VC approach was proposed to increase the computational efficiency of MRI image reconstruction for data sets acquired with a large number of receive coil elements [12,15,16]. Because techniques like GRAPPA reconstruct unaliased k-space observations for each coil, reducing the number of coils reduces the computational burden of the reconstruction problem by a factor nearly equal to the coil compression factor. With optimal coil compression, signal-to-noise ratio (SNR) or parallel imaging performance should not be compromised, while reducing the memory and computational footprint of image reconstruction. However, the appropriate coil compression factor depends on receive coil geometry and the amount of parallel imaging acceleration. Therefore, optimal coil compression factors must be determined for specific experimental conditions.

It is clear that the multitude of hyperparameters in SMS reconstruction can impact the quality of reconstructed images and downstream image analyses. This particularly affects neurocognitive imaging experiments that explore small-scale changes in voxel intensity over acquisitions comprising thousands of images. While the term *optimality* can have varying definitions, the goal of the current study was to increase the fidelity of SMS reconstruction with respect to a single-band reference image set. Altering reconstruction hyperparameters, such as estimated RF coil sensitivity profiles (actual or virtual), the number and location of calibration k-space lines, calibration kernel sizes and calibration k-space locations, have been shown to impact image quality [7–9]. Furthermore, it is well-known that use of coil compression significantly reduces computational time for image reconstruction, but it can also impact image quality. In neurocognitive experiments, it is posited that an optimal reconstruction algorithm for a given set of experimental parameters will yield images with minimal reconstruction artifacts that can be reconstructed with modest computational needs. In this work, a selection of hyperparameters for a hybrid-space coil-by-coil slice unaliasing approach [17] are utilized in reconstruction of fMRI time series data. The impact of a broad range of hyperparameter values on image quality are first analyzed through an exhaustive grid search. Subsequently, the impact of applying coil compression in combination with the optimized SMS hyperparameters is considered.

## 2. METHODS

Two experiments were performed in this study. The first experiment is designed to examine the relative impacts and interactions of hyperparameters. Such hyperparameters include the region of hybrid space used for unaliasing kernel calibration, the geometry of the hybrid space kernel used for unaliasing, the regularization of the kernel fitting process, and the inclusion or degree of channel compression performed prior to kernel fitting. Through this experiment, it was hypothesized that a unique set of hyperparameters would yield images with minimized artifacts.

The second experiment is designed to examine the impact of coil compression on reconstruction time and practical fMRI observations. All other reconstruction hyperparameters were fixed at their optimal values found in experiment 1. In both experiments, the receive coil used and the SMS acceleration factor were also varied to investigate the generalizability of the observed parameter interactions across varying experimental conditions. Details of the acquisitions, reconstructions, pre-processing, and analysis for these experiments are provided in the following subsections.

### 2.1. Data Acquisition:

This study was approved by the Institutional Review Board and prospective written consents were obtained from all participants. Experiments were conducted on a GE Healthcare Signa Premier 3.0T system (GE Healthcare, Waukesha WI), using both a 48-channel head receive coil (GE Healthcare, Waukesha WI) and a 32-channel head receive coil (Nova Medical, Wilmington MA). The 32-channel coil was smaller than the 48-channel with 23.5cm opening in the anterior/posterior (A/P) direction and 18.5cm in left/right (L/R) direction while the 48-channel coil was 25.7 cm in A/P direction and 23.0 cm in L/R direction. Data were acquired from five healthy volunteers (3 males, 2 females, four of them right-handed, 27±3 years of age, 70±15 kg). Blipped-CAIPI [5] was used for all SMS acquisitions. The acquisition parameters were chosen to best mimic those utilized in HCP acquisition protocols while still enabling the study's SMS performance analysis goals [1].

For the first experiment, twelve fMRI data sets were acquired from each participant using the two RF coils. A broad range of hyperparameters were examined using five resting state fMRI data acquired with SMS factors of 2, 3, 4, 6, and 8. A sixth series with no SMS acceleration was acquired and used as reference for quantitative comparisons. These acquisitions were repeated for both coils, resulting in twelve fMRI data sets. In order to facilitate the comparisons, all acquisition parameters were matched, with the exception of SMS factor. For all runs, a series of 114 volumes were acquired with TR=1100ms, TE=30ms, flip angle=90°, 24 slices with 4mm gap, an isotropic voxel resolution of 2mm and 20% partial-Fourier undersampling in the $k_y$ dimension. These settings provided consistent slice locations for all SMS factors tested and the slice gap of 4mm helped reduce impact from slice crosstalk effects due to imperfect slice profiles. Note that the minimum TR was not used for each of the SMS acquisitions. Rather, the minimum TR for the single-band experiment (1100ms) was used for all acquisitions in order to facilitate quantitative image comparisons in the study analysis. For SMS acquisitions, CAIPI field of view shifts were left unchanged from the vendor product paradigm.

For the second experiment, fMRI time series were acquired using a right-hand finger tapping stimulus and repeated for both coils. A single-shot gradient echo EPI was used for this acquisition. This data was used to study the impact of coil compression on fMRI acquisitions using a well-known and robust fMRI paradigm. Total acquisition time was 4 minutes and 20 seconds with TR=800ms, TE=30ms, flip angle=50° and an SMS factor of 8. 72 contiguous axial slices were acquired in an interleaved fashion with 2mm isotropic resolution. Tasks were presented using a block design, which consisted of 7 resting blocks interspersed with 6 finger tapping periods, each 20s long. Participants were visually cued to start and stop and instructed to tap their fingers approximately once per second. Images were reconstructed with a kernel size of 7*2 ($x$ and $k_y$), calibration area size of 32*96 ($x$ and $k_y$) and regularization weight of $10^{-2}$. These parameters seemed to work consistently well for the range of VCs tested with SMS factor of 8 for the two coils.

A calibration scan was acquired in order to calculate reconstruction kernels for all scans with SMS acceleration. This calibration scan was acquired with a 3D SPGR pulse sequence with TR=1.4ms, TE=0.5ms, 30cm FOV, 9.4mm in-plane resolution, 10mm slice thickness and 32 slices. Images were acquired with the body coil and the respective receive coil to generate coil sensitivity maps. A T1-weighted image was also acquired from each participant for anatomical reference using MPRAGE with isotropic resolution of 0.8mm, inversion time of 1060ms and TR=5.4ms and TE=2.1ms.

**2.2. Image Reconstruction:**

Image reconstruction was performed offline with locally modified software which included code from a vendor-supplied externally calibrated SMS reconstruction algorithm (slice-ARC, GE Healthcare, Waukesha, WI). Software modifications were performed utilizing a C++ application programming interface provided by the vendor (Orchestra SDK, GE Healthcare, Waukesha WI). Library functions for coil compression, slice-ARC SMS, Fourier reconstruction and coil combination were used.

Acquired data were retrospectively reconstructed with varying hyperparameters. Since multi-parametric optimization with the hyperparameters of interest would be computationally expensive for the first experiment, a heuristic grid search was conducted by testing a range of values for each hyperparameter within a practical search space. The sizes of slice-ARC kernels were 5, 7 or 9 in the $x$ direction, and 2, 4, 6 or 8 in the $k_y$ direction. Similarly, the size of calibration area was 16, 32 or 48 in the $x$ direction and 96, 144 or 192 in the $k_y$ direction. Tikhonov regularization was used for kernel fitting and the regularization weights of $10^{-2}$, $10^{-4}$ and $10^{-6}$ were tested.

Coil compression using singular value decomposition at each location was performed using library functions in the Orchestra SDK that implemented the method published previously [16]. In this method, explicit derivation of coil sensitivity profiles or estimation of noise correlation matrices were not needed. VC channels were calculated using the external calibration scans acquired for slice-ARC. The minimum number of VCs was set to 8 for both coils and then incremented by 8 up to a maximum of 32 VCs with the 32-channel coil and up to 48 VCs with the 48-channel coil. Slice-ARC kernel weights were estimated for each VC set.

As noted earlier, a grid search across this full set of parameters was performed on the five acquired SMS datasets for each subject and RF coil. For each SMS acquisition, and each subject with the 32-channel coil, 1620 unique time series were reconstructed. Because of an increased range of reconstructed virtual channels including 40 and 48, 2268 unique time series were reconstructed with the 48-channel coil. Across the five imaged participants and five separate SMS acceleration factors, 97,200 unique time series were thus reconstructed for the first experiment. For the second experiment that utilized a task-based fMRI acquisition, all hyperparameters except the number of virtual coils were held constant at their optimal values. First, an aggressive coil compression was applied to both RF coils, yielding 8 virtual channels. Then, 50% coil compression was tested, resulting in 16 and 24 channels respectively for the 32- and 48-channel coils. Data were also reconstructed with no coil compression for comparison.

### 2.3. fMRI data processing and analysis:

All fMRI data were preprocessed using pipelines developed in FSL software package, version 6.00 [18,19]. Prior to data analysis, each fMRI run was corrected for head motion using rigid body affine registration (mcFLIRT).

#### *2.3.1. Analysis of slice-ARC hyperparameters on reconstructed image quality:*

The six task-free fMRI time series acquired for each subject and coil were used to evaluate slice-ARC performance across varying hyperparameters. Single-band image reconstructions from the 32-channel RF coil yielded the highest temporal signal-to-noise ratio (tSNR) and were thus selected as the ground truth. The quality of an SMS image was evaluated by calculating the correlation between the single-band reference image volume and the SMS image volume reconstructed with a given set of hyperparameters. Since the same image contrast and resolution is maintained across all acquisitions with and without SMS, a high correlation between unaliased SMS image volumes and single-band image volumes is indicative of effective image unaliasing.

#### *2.3.2. Analysis of fMRI signal detection performance with coil compression:*

Brain activation statistics for finger tapping fMRI data were computed using FEAT software tool [20] in FSL software package. Voxel time series were temporally filtered using a high pass filter with a 100s cut-off to eliminate low frequency fluctuations and temporal drifts, and spatially smoothed using a 4mm full-width-half-maximum Gaussian kernel. A General Linear Model for ANOVA with one factor, three levels with repeated measures was setup for this analysis. For the 32-channe coil, the three levels of VCs chosen were 8, 16 (50% compression) and no compression. For the 48-channel coil, the VCs compared were 8, 24 (50% compression) and no compression.

Each subject's data was analyzed in native space and then statistic maps were registered to standard space for group analysis. A two-step registration procedure was used to transform statistic maps to the standard space. The T1 weighted images were registered to the Montreal Neurological Institute MNI-152-T1 standard space brain image using 12 parameter affine transformations using FLIRT software tool [21,22] followed by FNIRT nonlinear registration [18,19]. The statistic images were first registered to the T1 image and then the same transformations were applied to transform them to the stereotaxic space. This approach enabled the group level analysis of the impact of coil compression on the statistical brain activation maps.

## 3. RESULTS
### 3.1. Analysis of SMS hyperparameters on reconstructed image quality:

Image reconstruction performance varied greatly across the range of hyperparameters considered in this grid search. The full range of the correlations between unaliased SMS image volume and single-band image volume was 0.67 to 0.98, with a median of 0.86. Hyperparameters for the $x$ direction calibration region and the $x$ direction kernel size were found to have no observable impact on unaliasing performance.

With the lack of performance dependence on $x$-direction calibration region sizes and $x$-direction kernel sizes and consistent results across subjects, correlation data were reduced through a median operation across those variables for each remaining combination of hyperparameters. The combined effect of five remaining hyperparameters ($k_y$ calibration region size, $k_y$ kernel size, regularization, coil compression, and RF coil) on the quality of images were illustrated on radar plots. A guide to the generated radar plots is provided in Figure 1 to facilitate interpretation. In all the radar plots, the radius is the correlation coefficient, shown on a power scale, such that outer rings show higher correlation. Different symbol shapes represent the kernel size in the $k_y$ direction, and the symbol size represents the calibration area size in the $k_y$ direction. The color of the plotted symbols corresponds to the 10-base logarithm of the Tikhonov regularization term, as shown with the colorbar. A subset of representative unaliased SMS images with varying levels of correlation are shown with their markers in Figure 1.

Figures 2 and 3 show the performance of image reconstruction with different hyperparameter combinations for the 32-channel and 48-channel RF coils, respectively. Note that the radius for plots is scaled on a power scale between 0.79 and 0.99, and between 0.72 and 0.90 on the 32-channel and 48-channel RF coils, respectively, to show differences in better detail.

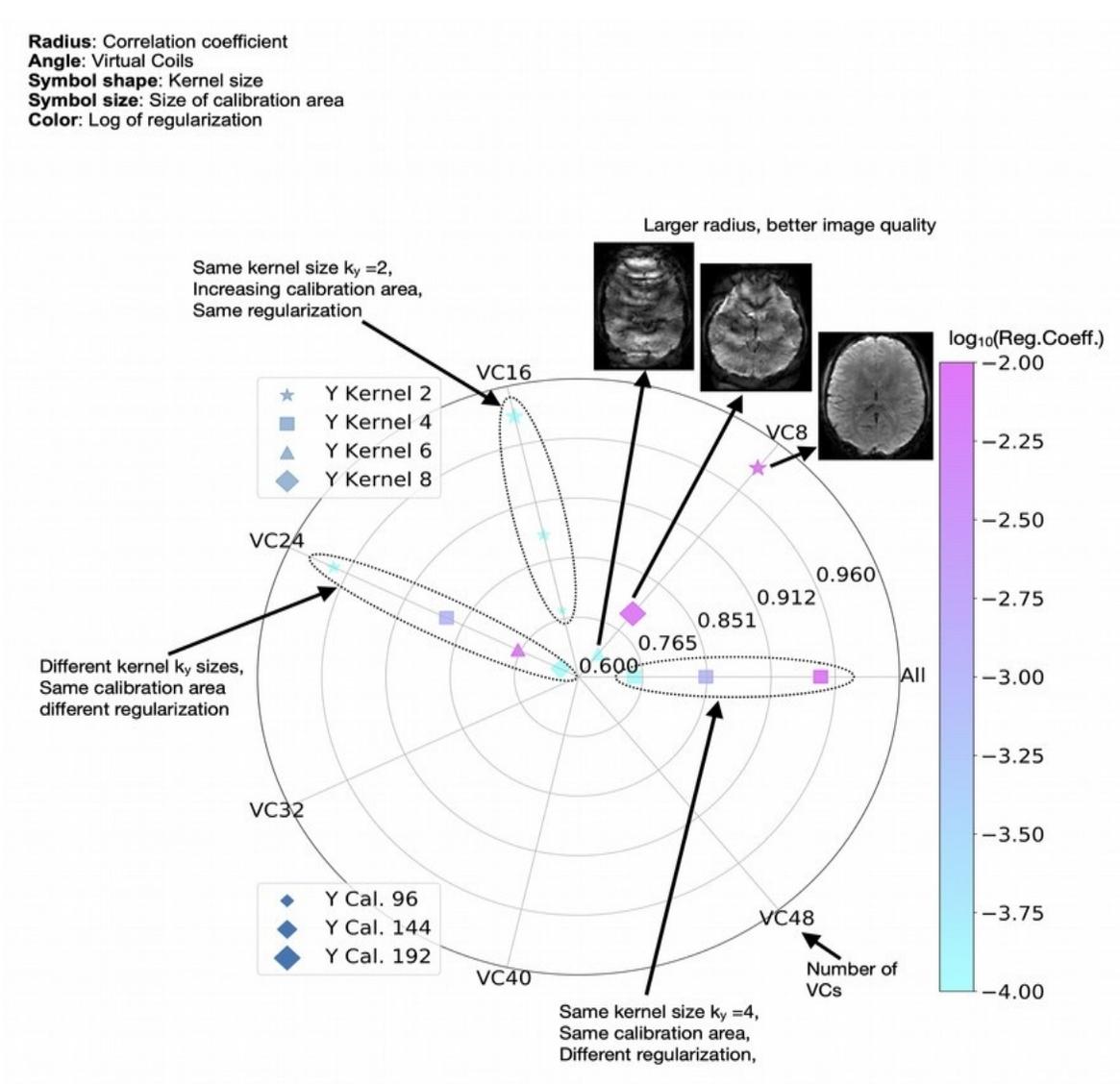

Figure 1. An illustrated guide to radar plots.

### 3.1.1. Selection of Coil:

In all cases considered in this analysis, the 32-channel coil shows higher correlation between the unaliased SMS images and single-band images compared to the 48-channel coil. Different radial scales are shown with radar plots for the 32-channel and 48-channel coils, because maximal correlations for the 32-channel coil were greater than 0.95 for each SMS acceleration factor, while maximal correlations for the 48-channel coil were less than 0.90 for all acceleration factors. This is consistent with the fMRI analysis of tSNR (data not shown) in which the performance of the 32-channel coil was also found to be superior. In general, the impact of coil compression and SMS acceleration factors up to 6 were relatively minor on the unaliasing performance of the 48-channel coil, provided optimal size of kernel, calibration area and regularization were employed.

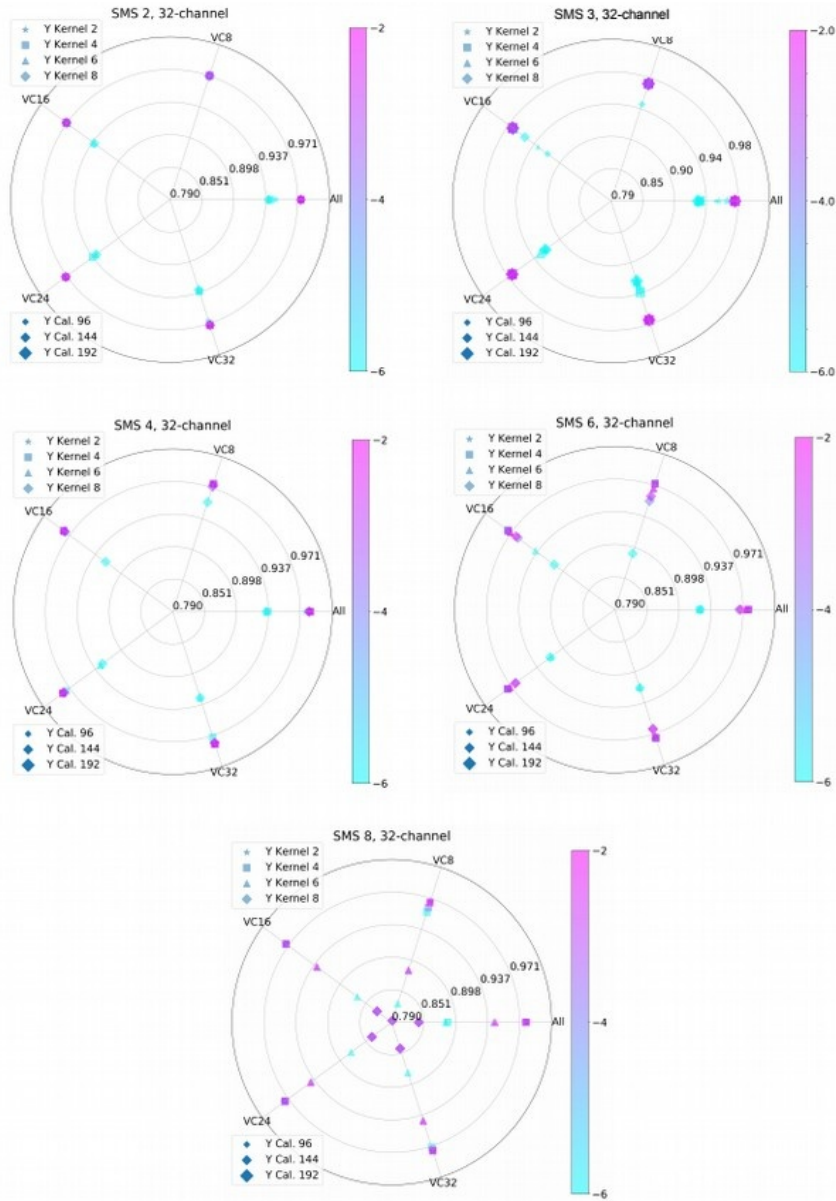

Figure 2. Performance in the 32-channel RF coil with changes in hyperparameters

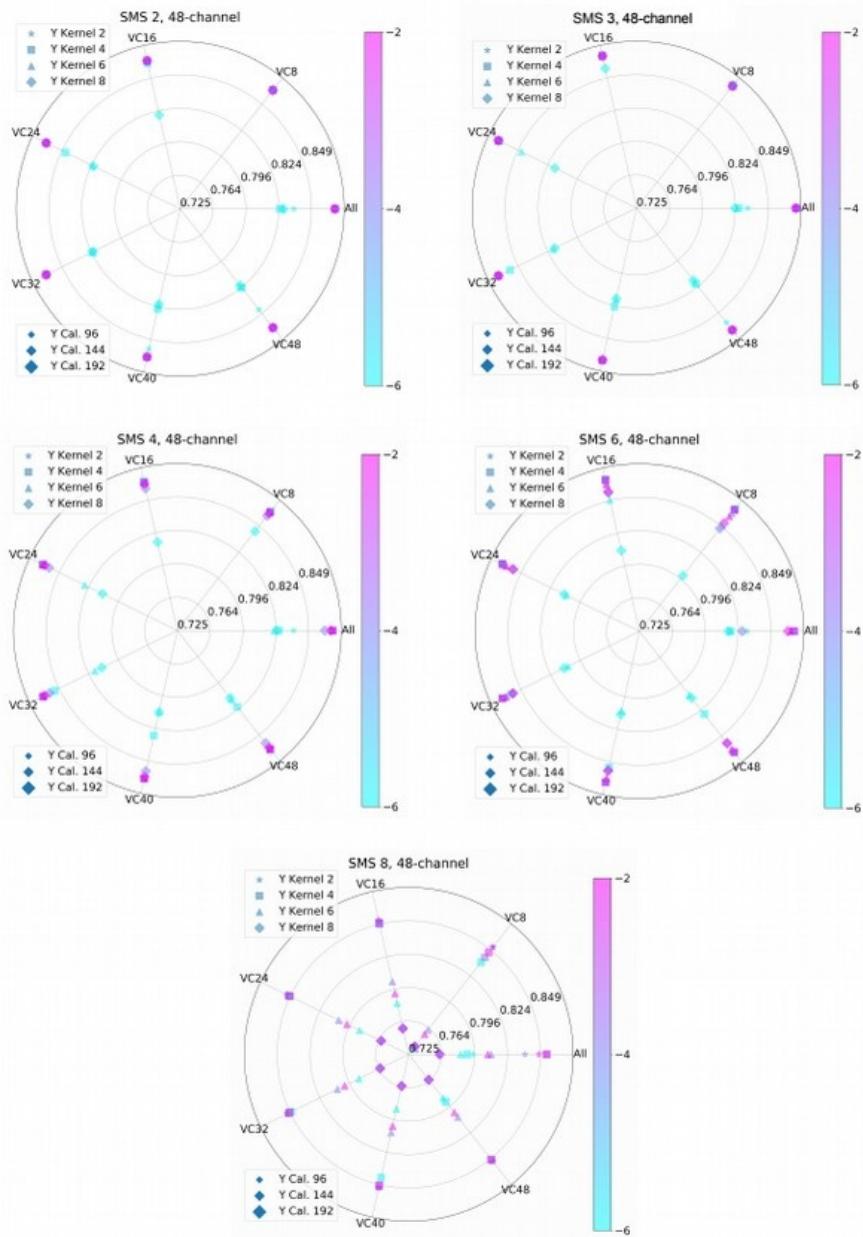

Figure 3. Performance in the 48-channel RF coil with changes in hyperparameters

### *3.1.2. Regularization:*

Increased regularization improved correlation between unaliased SMS images and reference single-band images. Across acceleration factors, calibration regions, kernels, and RF receiver coils tested, Tikhonov regularization factors of $10^{-6}$ performed the worst in general. However, there were exceptions. For example, with SMS 8, VC 16 or VC 24 using the 48-channel coil had better outcome if a small kernel with light regularization were used. In the 32-channel coil, higher regularization between $10^{-4}$ – $10^{-2}$ generally improved image quality for all SMS factors and VCs. But, heavier regularization of $10^{-2}$ slightly decreased performance compared to $10^{-4}$, even when other hyperparameters with optimal values were selected. Similarly, when smaller kernels were used, increased regularization was found to improve unaliasing performance, although there were exceptions.

### 3.1.3. $k_Y$-Calibration Region Size:

The size of the calibration region in the $k_Y$-direction influences the correlation between unaliased SMS data and single-band images. Generally, increased calibration region size yielded increased correlation. The improvement from a calibration region size of 96 to 144 points is much greater than the improvement from a region of 144 points to 192 points. As mentioned in the results associated with regularization, increased calibration regions tend to yield improved results with increased regularization.

### 3.1.4. $k_Y$-Kernel Size:

The impact of $k_Y$-kernel size and unaliasing performance is marked. Large kernel sizes, including 6 and 8 points, yielded reduced correlation between unaliased SMS images and single-band images. Smaller kernel sizes of 2 and 4 performed equally well, when other hyperparameters were selected accordingly. The impact of regularization was greater with smaller kernels, compared to larger kernels.

### 3.1.5. Coil Compression:

With optimal selection of other hyperparameters in the utilized SMS unaliasing algorithm, coil compression caused very slight reduction in correlation between single-band and SMS unaliased images across the tested range of SMS acceleration factors and the two tested RF coils. When small regularization factors, smaller calibration regions, and poorly performing kernels are selected, the impact of extreme coil compression, including use of a low number of virtual channels with a high SMS acceleration factor, yields a more marked degradation in unaliasing performance.

### 3.1.6. SMS Acceleration:

Unaliasing performance was found to vary with the SMS acceleration factor. The SMS acceleration factor of 4 had the best performance overall compared to all other acceleration factors. The second best results were with SMS acceleration factor of 6, which outperformed SMS factors of 2 or 3. This counter-intuitive observation is further discussed in the next section. Up to acceleration factors of 4, the primary driver of unaliasing performance is the regularization factor. In those cases, if the regularization factor is greater than or equal to $10^{-4}$, unaliasing performance is not grossly impacted by the selection of other SMS hyperparameters. In the regime of high factors of SMS acceleration, the variance of unaliasing performance with hyperparameter selection becomes clear. In this regime, optimal unaliasing performance is achieved only when each hyperparameter is tuned properly.

### *3.1.7. Hyperparameter Interactions:*

While, in most cases, good performance can be achieved by selecting the high-performing hyperparameters described above, hyperparameter combinations yielding best performance were found to not always be the simple combination of the best, individually considered parameters. Typically, interaction effects between regularization levels and other hyperparameters became more pronounced with higher SMS acceleration factors of 6 and 8. In some cases, large calibration area required heavier regularization. An example of this can be seen in the 32-channel coil with SMS acceleration of 4 with kernel size of 6, calibration size of 192 and 24 VCs, which needed heavier regularization to converge. Kernel size also had a significant influence on the slice-ARC performance, and optimal size depended on other hyperparameters. For example, with the 32-channel coil, kernel size of 2 did not perform well with no coil compression and low regularization but produced good results for SMS factor of 6. On the other hand, kernel size of 8 performed poorly for SMS factor of 6, and kernel size of 4 was the best performer for SMS factor of 8. In fact, $k_Y$-kernel size demonstrated strong interactions with other reconstruction hyperparameters.

In order to illustrate interactions between hyperparameters, three interesting cases are selected from the 32-channel RF coil and illustrated in Figure 4. In these plots, three of the five hyperparameters are fixed and the other two are varied to show how those two influence each other. The first plot (a) shows how the correlation between the reference image and SMS image changes with different $k_y$-kernel sizes as $k_y$-calibration size changes. Similarly, the plots in (b) shows how different regularization factors affect image quality as $k_y$-kernel size changes. The last panel (c) illustrates the impact on image quality with different VCs and $k_y$-kernel sizes.

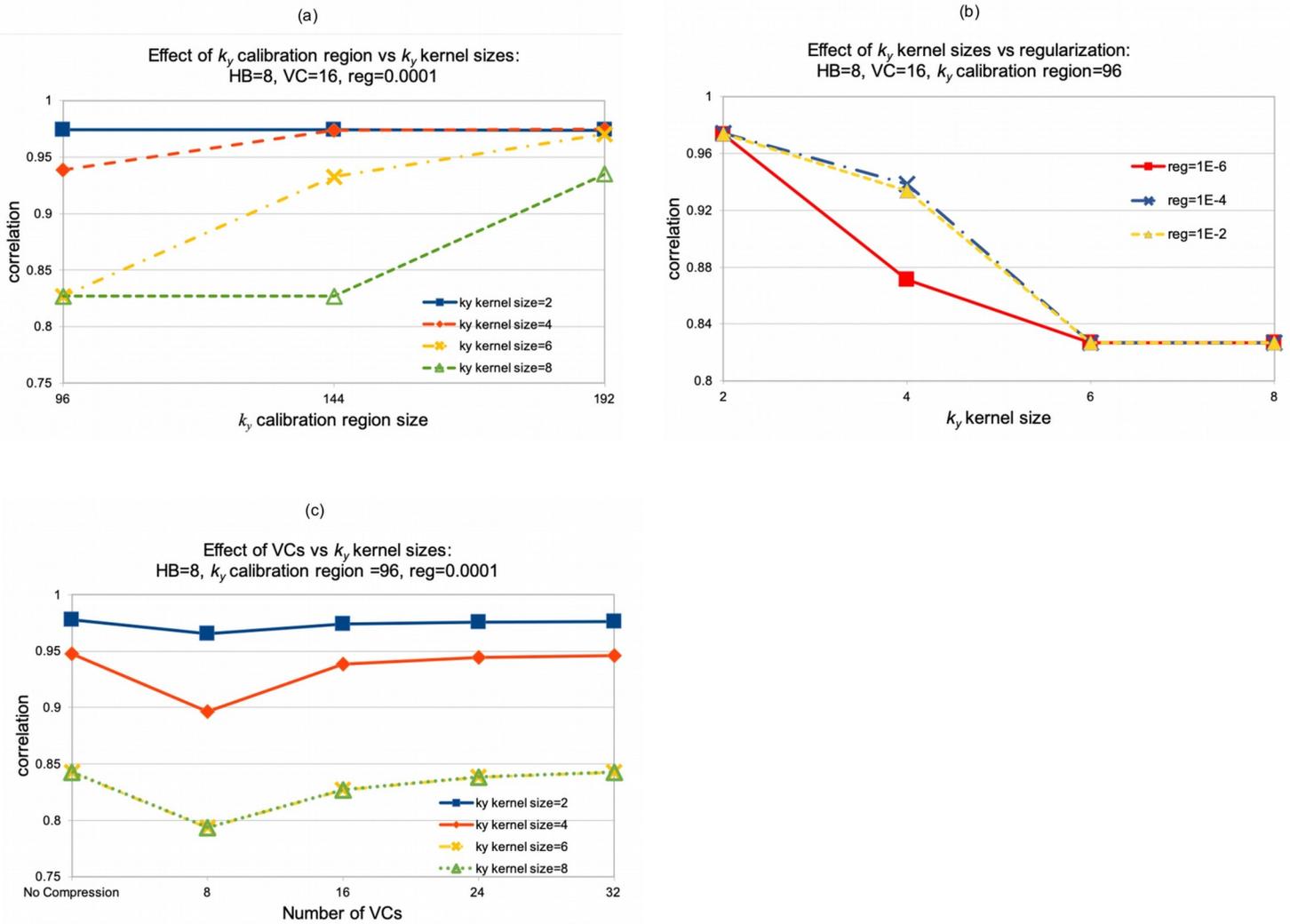

Figure 4. Interactions between hyperparameters for the 32-channel RF coil. For each plot, three of the five hyperparameters were fixed and the other two were varied. (a) Changes in correlation between the reference image and SMS image with different $k_y$-kernel sizes as $k_y$-calibration region increases. (b) Impact on image quality with different regularization factors and $k_y$-kernel sizes. (c) Impact on image quality with different VCs and $k_y$-kernel sizes.

3.2. **Analysis of fMRI signal quality with coil compression:**

Figure 5 shows group analysis of fMRI finger tapping experiments. Results with 50% compression and no compression are displayed for the two RF coils for qualitative comparison. Statistical parameter maps are shown for p<0.05, corrected for multiple comparisons using False Discovery Rate (FDR) [23].

Results of ANOVA analysis with coil compression as factor are illustrated in Figure 6. The comparisons did not survive p<0.05, with FDR correction. However, there were notable trends without multiple comparison correction. Therefore, statistical maps are shown for p<0.01, uncorrected. Post-hoc analysis for comparing each coil compression level against no compression showed some differences where blood oxygen level dependent (BOLD) effects in images with no coil compression were slightly higher than those with 8 or 24 VCs for the 48-channel RF coil. Moreover, the results with 8 VCs had more clusters with differences than those with 24 VCs. On the other hand, no notable differences were seen in the 32-channel RF coil, even without FDR correction.   There was only a very small cluster when compared with 8 VCs. Despite the high SMS acceleration factor of 8, the fMRI signal was still robust with just 8 VCs from the 32-channel coil.

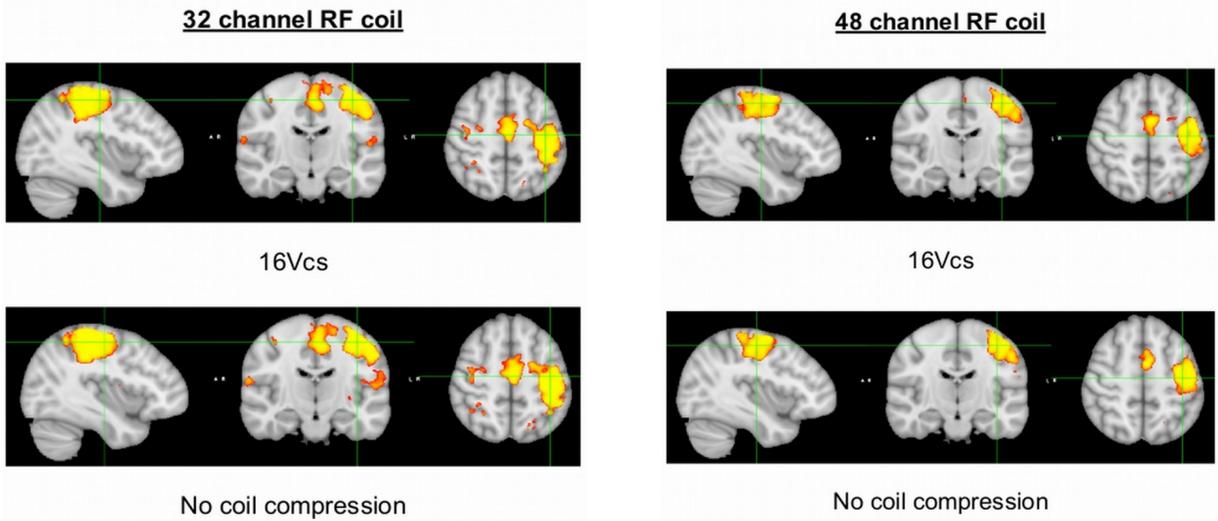

Figure 5. Group fMRI results with no coil compression and with 16 VCs for qualitative comparisons. Statistical parameter maps are shown for p<0.05, FDR corrected.

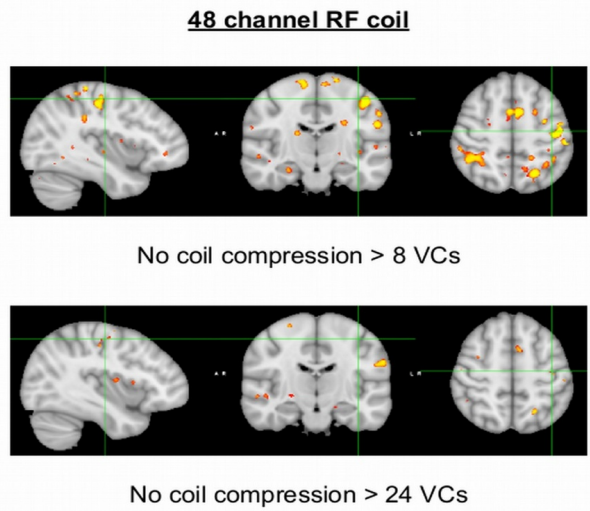

Figure 6. Results of ANOVA analysis with coil compression as the factor with repeated measures from five subjects. Results are shown for p<0.01, uncorrected.

Finally, the computation time for different coil compression rates were plotted in Figure 7 for each SMS acceleration factor. CPU time increased linearly with increasing VCs.

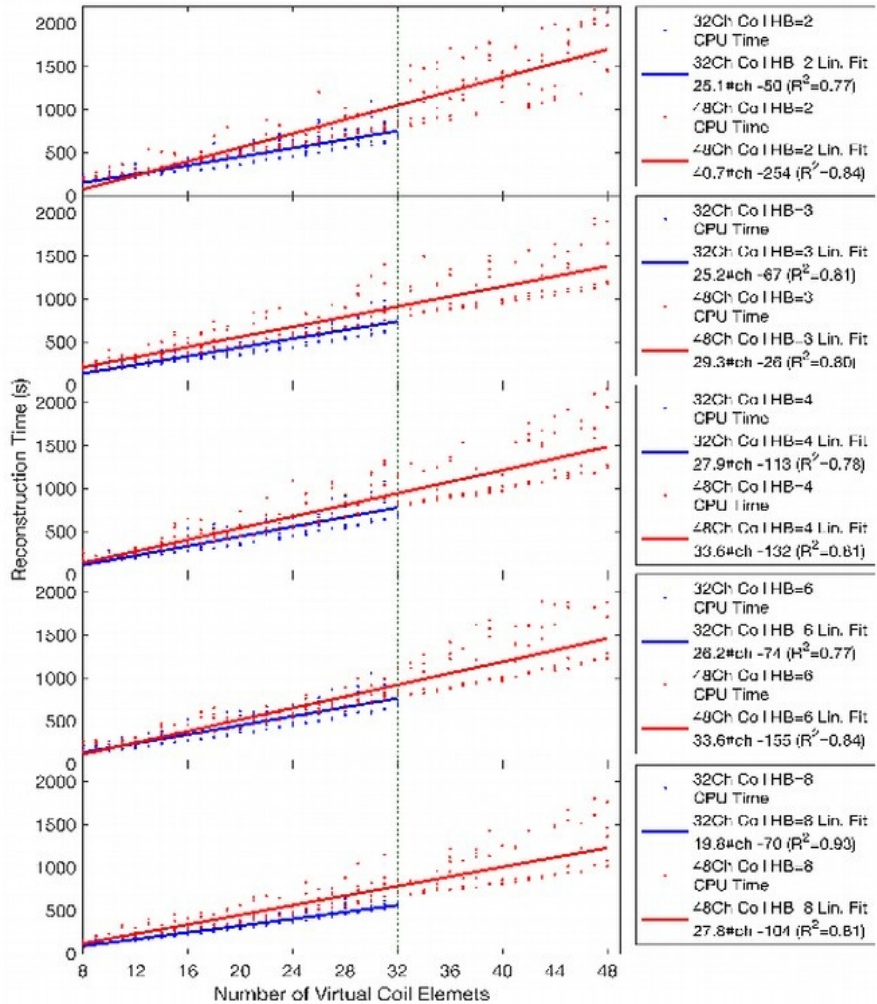

Figure 7. CPU time with respect to coil compression for the 32-channel coil (blue dots) and 48-channel coil (red dots) for the five subjects. The results with increasing SMS factors are shown in each consecutive row. Solid lines show the best fit with linear regression. Linear regression coefficients and goodness of fit, $R^2$, are listed in the legends. For all regressions $p<0.0001$, and F statistics > 20.

## 4. DISCUSSION

SMS technologies are commonly used in fMRI studies that require high spatial and temporal resolutions, including acquisitions harmonized with HCP [1]. This study explored the variation of SMS unaliasing performance as a function of SMS acceleration rates and reconstruction parameters for a particular commercially available experimental configuration of single-shot gradient-echo EPI SMS. Acquisitions were performed with vendor-configured CAIPI shifts and

SMS reconstructions utilized a vendor-supplied externally-calibrated hybrid-space unaliasing algorithm. Two RF receive coil arrays with different channel counts and dimensions were also evaluated. The results showed clear trends indicating that hyperparameter settings such as regularization level and calibration kernel size, can impact reconstruction performance. These parameters exhibited coupled dependencies and varied with different experimental setups, such as RF coil used, acceleration rate and coil compression factor.

Between the two RF coils tested, the 32-channel coil with reduced form factor clearly outperformed the larger 48-channel coil in our experiments. This is probably due to the closer proximity of the coil elements to subjects' heads in the 32-channel coil. SMS acceleration factor of 4 outperformed all other acceleration rates, and SMS acceleration of 6 showed only slight drop in performance in the 32-channel coil. This somewhat counterintuitive result was probably due to suboptimal g-factors for 2- and 3-fold SMS accelerations arising from the interaction of the RF coil sensitivity profiles and selected CAIPI field of view shifts (g-factor is a term that describes the noise amplification in parallel MRI methods [3,24]). Note that the CAIPI shifts utilized were hard-coded in the vendor acquisition, and they were not considered as an optimization hyperparameter in this study. However, the hypothesized interaction between coil geometries, CAIPI shifts, and SMS performance that has been revealed in this study warrants further investigation.

For both coils, coil compression did not result in significant reduction in image quality up to 16 VCs. However, there was noticeable reduction in data quality when more aggressive compression with 8 VCs were utilized. This was supported by the finger-tapping fMRI results shown in Figures 5 and 6. While 16 VCs yielded fMRI activation maps that were not statistically different from those with no-compression images, 8 VCs showed noticeable differences. This effect was again more pronounced in the 48-channel coil. On the other hand, image reconstruction was typically ~3 times faster with 16 VCs. That is a considerable improvement in image reconstruction speed with negligible penalty in image quality, especially when coil arrays with large number of elements were used and a long time series of images are acquired. However, one needs to tune the other image reconstruction parameters to make the best use of coil compression.

One of the parameters that had a big influence on image quality was the regularization weight that was used in the data fitting for estimating SMS kernel coefficients. A regularization weight of about $10^{-4}$ appeared to be a proper choice for the tests conducted in this study. Both lower and higher regularization typically resulted in lower image quality. This is in line with the literature where different approaches were suggested to find the optimal regularization [10]. Regularization penalizes large estimates for kernel coefficients so that a few measurements in calibration data do not have too much influence on reconstructed image. If the applied regularization is too low, it will still allow such large coefficients. On the other hand, too much regularization smooths out the kernel and does not allow required variation in the computed kernel for accurate reconstruction.

Another pair of important observations from this study pertain to the kernel and calibration region sizes in the $k_y$ dimension. The observed trends (improved performance with smaller kernel sizes and larger calibration region) may be related to the specific vendor-provided SMS paradigm applied in this study, wherein a non-EPI (low resolution 3D-SPGR) based calibration scan with a flat contrast is utilized for kernel fitting. When utilizing a smoothly varying image set for calibration, improved calibration performance using compact convolution kernel in k-space is expected due to the Fourier convolution theorem. The improved performance with the increased size of the calibration region may be related to the interplay between kernel fitting performance and the presence of outliers in the calibration data. Such outliers tend to cluster in the lower spatial frequencies and are, as a result, over-represented in calibration data with a smooth calibration image and a small calibration region [25]. Interestingly, variations of applied kernel parameters in the frequency encoding ($x$) dimension had minimal effect on SMS performance. This observation may be due to interactions of the hybrid space SMS approach and low resolution external calibration utilized in the vendor-provided workflow.

The results from this study can be used to provide guidance to researchers applying the widely available vendor-provided SMS workflow utilized in this study (external 3D-SPGR calibration, hybrid-space SMS algorithm, and prescribed CAIPI-shift paradigm). Specifically, for high resolution axial gradient echo EPI acquisitions that mimic the settings utilized in HCP acquisition protocols [1], the following hyperparameters would be recommended: $k_y$ kernel size of 2, $k_y$ calibration size of 192, Tikhonov regularization of $10^{-4}$, and 24 virtual coils. These recommendations are largely independent of SMS acceleration factors or either of the head array coils analyzed in this study.

## 5. CONCLUSIONS

In this work, a grid search of hyperparameters for a vendor-provided hybrid-space SMS algorithm was conducted to identify best practices for SMS reconstruction hyperparameter selection. In general, $x$-dimension variables offered little impact, while larger calibration regions and moderate to small kernel sizes in $k_y$ direction, SMS acceleration factors less than 6, and the use of a 32-channel RF coil yielded better results. This study demonstrated the feasibility and anticipated performance of advanced functional imaging studies with high SMS factors using two RF coils approved for clinical imaging. The methods utilized in this study deliberately relied upon commercially available acquisition and reconstruction methods. A key takeaway from the study is that the hyperparameters used in SMS reconstruction need to be fine-tuned once the experimental factors such as the RF receive coil and SMS acceleration have been determined. More generally, a cursory evaluation of SMS acquisition and reconstruction hyperparameter values is recommended before conducting a full-scale study utilizing a given experimental setup. Based on the experiments performed in this analysis, a recommended set of hyperparameters has been provided for acquisitions following the HCP protocol. However, it must be emphasized that different experimental parameters, such as imaging planes, image resolution, SMS unaliasing algorithms, RF coils, CAIPI shifting strategies, and coil compression algorithms will require additional analyses to derive appropriate hyperparameter recommendations.

**ACKNOWLEDGEMENTS**

This work is supported by GE Healthcare technological development grant and Daniel M. Soref Charitable Foundation. We would like to thank Dr. Alexander Cohen for his help on fMRI processing and acknowledge the support of the MCW Research Computing Center for computational resources.